# Hearables: Multimodal physiological in-ear sensing

Valentin Goverdovsky[a], Wilhelm von Rosenberg[a], Takashi Nakamura[a], David Looney[a], David J. Sharp[b], Christos Papavassiliou[a], Mary J. Morrell[c], and Danilo P. Mandic[a,1]

[a]Department of Electrical and Electronic Engineering, Imperial College London, SW7 2BT, United Kingdom; [b]Computational, Cognitive, and Clinical Neuroimaging Laboratory, Centre for Neuroscience, Division or Brain Sciences, Imperial College London, W12 0NN, United Kingdom; [c]Academic Unit of Sleep and Ventilation, National Heart and Lung Institute, Imperial College London, NIHR Respiratory Disease Biomedical Research Unit at the Royal Brompton and Harefield NHS Foundation Trust and Imperial College, SW3 6NP, United Kingdom

This manuscript was compiled on September 12, 2016

**Future health systems require the means to assess and track the neural and physiological function of a user over long periods of time and in the community. Human body responses are manifested through multiple modalities, such as the mechanical, electrical and chemical; yet current physiological monitors (actigraphy, heart rate) largely lack in both the desired cross-modal and non-stigmatizing aspects. We address these challenges through an *inconspicuous and comfortable* earpiece, equipped with miniature multimodal sensors, which benefits from the relatively stable position of the ear canal with respect to vital organs to robustly measure the brain, cardiac and respiratory functions. Comprehensive experiments validate each modality within the proposed earpiece, while its potential in health monitoring is illustrated through case studies. We further demonstrate how combining data from multiple sensors within such an integrated wearable device improves both the accuracy of measurements and the ability to deal with artifacts in real-life scenarios.**

wearables | physiological sensing | in-ear sensing | multimodal sensor

Recent advances in wearable technology and the Internet-of-Things have opened the possibility for monitoring human physiological functions out of the clinic and in the community. Once fully developed, this is envisaged to significantly alter the landscape of healthcare through continuous 24/7 management of diagnosis and treatment. Early attempts in this direction employed actigraphy, which uses inertial sensors to measure activity of various body parts (BodyMedia), with applications in sports science and general health. More recent efforts have focused on recording cardiac activity, either electrically, via electrocardiograms (ECG), or optically, through photoplethysmograms (PPG). These require either a chest-strap or wrist-band, with the latter considered inconspicuous and comfortable to wear for most people, and the former quite cumbersome.

The development of wearable devices for other modalities, such as respiration and neural activity, has been largely hampered by the inconvenience of their respective form factors – a chest-strap within respiration monitors is uncomfortable for long-term use, while the electroencephalogram (EEG) recorded from the scalp is cumbersome to set up, stigmatizing when out-of-the-clinic, and prone to artifacts. More recent technologies, such as AcuPebble, make use of very sensitive miniature microphones integrated within a wireless device placed over the suprasternal notch, allowing the detection of minute sounds produced by the turbulent airflow in the lungs, as well as heart beats. Such devices are envisaged to find application in the monitoring of sleep apnea, chronic obstructive pulmonary disease, and asthma.

The de facto standard for measuring brain electrical activity is electroencephalography which involves the placement of a *multielectrode array* onto the scalp and requires an electrolyte to enhance the contact between the electrodes and the skin. Recent efforts to improve this technology have focused on reducing the number of electrodes, exploration of new materials (composites [1, 2], nanowires [3]) and their alternative physical forms (spikes [4], needles [5, 6], bristles [7], capacitive disks [8]), together with producing visually appealing designs (Emotiv). Despite continuous improvements in the areas of electrode design and user-friendliness, the widespread out-of-the-clinic use of head-mounted EEG-devices is unlikely, due to their obtrusive, stigmatizing, and non-discreet nature.

One novel solution for continuous and unobtrusive EEG monitoring uses a curved electrode system integrated on the auricle, with the electrodes operational over two weeks, provided the spray-on bandage is regularly reapplied to the device [9]. The authors state that the electrode placement procedure – critical to ensure high EEG quality – is non-trivial. Furthermore, new form factors – in-ear [10, 11] and around-the-ear [12] – have recently emerged which allow the electrodes to be concealed in an inconspicuous EEG acquisition device.

Despite recent advances, the state-of-the-art health monitoring solutions rarely consider more than a single sensing modality. Yet, future health systems require that the whole spectrum of physiological responses is obtained with as few individual devices as possible, in order to both provide user convenience and gain aditional information through cross-modal coupling. We here introduce such a device, with the form factor of a standard in-ear headphone, routinely worn worldwide, which integrates miniature sensors for continuously monitoring neural, cardiac and respiratory activity from in-

**Significance Statement**

Wearable technology is envisaged to revolutionize the way we approach health and well-being, but for this to become a reality, a multitude of physiological sensors should be integrated into a single, comfortable-to-wear, unobtrusive and non-stigmatizing package. To this end, we introduce a multimodal system capable of monitoring cardiac, respiratory and neural activity from a set of miniature electro-mechanical and acoustic sensors integrated onto an earplug and placed inside the ear canal. The device is similar in its form factor to in-ear headphones, while the cross-modal information is shown to both enhance signal quality and facilitate artifact removal – key requirements in continuous 24/7 monitoring in the community.

side the ear canal. It is mechanically stable, comfortable to wear, unobtrusive, discreet, and straightforward to apply, and, with the advancement of system-on-chip technology, promises widespread continuous physiological monitoring in the community, leading to improved health management.

### Results and Discussion

**Sensor construction.** The proposed multimodal in-ear sensor comprises five key components: memory-foam substrate, two miniature microphones and two conductive cloth electrodes; Fig. 1*A* shows details of the device construction. The substrate material is a viscoelastic foam which allows for the absorption of artifacts stemming from both small and large mechanical deformations to the ear canal walls; these may arise from cardiac activity as well as chewing, swallowing and speaking. Furthermore, this flexible earpiece is generic, in the sense that it can be squeezed and shaped to fit any ear. When placed in the ear canal, memory foam relaxes and evenly redistributes the outward pressure throughout its outer surface, thus ensuring a comfortable and snug fit. Fig. 2*A* and 2*C* compare the viscoelasticity of different memory foams with silicone buds – another popular material choice for earplugs. The strain–stress curves were produced for 16 mm tall cylinders of both materials, which were first rapidly (10 mm in 2.4 sec) compressed in a dynamic mechanical analyzer; then, the strain was held constant for 1 min and finally gradually released at 2 mm/min rate. The two key points to observe are: 1) a dramatic relaxation of the dense foam when held for 1 min at 60% compression (from >80 kPa to <20 kPa), and 2) almost an order of magnitude lower compressive stress in memory foam compared to silicone (kPa vs MPa). These two desirable characteristics ensure comfortable fit and straightforward insertion of the earpieces based on memory foam, together with reduced mechanical artifacts from pulsatile ear canal movements due to blood vessel pulsation, as shown in Fig. 2*B* and Fig. 2*D*.

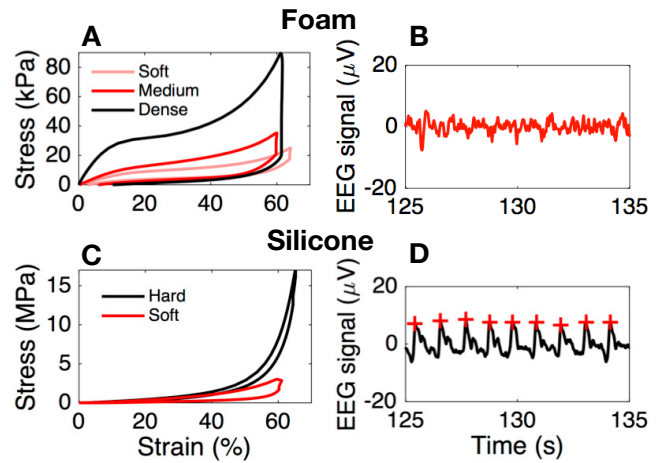

**Fig. 2.** Mechanical tests comparing the viscoelastic foam substrate and silicone. (A) Stress-Strain curve for viscoelastic foam cylinders of varying density. (B) EEG signal captured with the foam-based earplug from the ear canal of a person with strong blood vessel pulsation, observe no pulsation in the recorded EEG. (C) Stress-Strain curve for silicone cylinders of varying hardness. (D) EEG signal captured using the silicone-based earplug from the ear canal of a person with strong blood vessel pulsation, observe strong contamination of the EEG with pulsatile artifacts.

Electret condenser microphones (ECMs) are integrated within the viscoelastic earpiece to provide two sensing modalities: 1) mechanical disturbance measurements within the multimodal sensor (MMS) in order to deal with motion artifacts, and 2) recording speech and breathing activity from inside the ear canal. The MMS is constructed by first embedding an ECM within the body of the memory foam substrate such that it is aligned with the surface of the earplug, as shown in Fig. 1*B*. The microphone is subsequently sealed with a thin layer of plastic film. Finally, a cloth electrode is placed over the microphone and attached to the substrate with an adhesive, thus forming a flexible electro-mechanical electrode [13, 14]; the contralateral electrode is attached to the earpiece in the same way but with no ECM underneath.

Flexible cloth EEG electrodes are constructed out of conductive stretchable knitted fabric with less than $2\Omega/\square$ of surface resistivity. Fabric is cut into strips of 4×7 mm and all sensors on the multimodal earpiece are connected to signal acquisition equipment.

**Physics behind ear-EEG.** The physiology of neural activity measurements via scalp EEG is relatively well understood. In order to demonstrate that EEG can also be measured from inside the ear canal, a number of physics simulations were conducted in COMSOL Multiphysics® using realistic head models generated from MRIs. Fig. 3*A, B, C* show the brain electric potentials at time 0 (with respect to the phase onsets) on the head surface, created by the superposition of the electric fields produced by six current dipoles – two in *each* auditory cortex and two in the brain stem – which simulate the brain activity resulting in the auditory steady-state response (ASSR). The upper part of the head is positively charged (red) while the lower part is charged negatively (blue); for quantitative analysis, locations of the electrodes were chosen to be inside the ear canals, root of the helix, and at the standard scalp positions Cz and T8. Although the exact power of the acquired potentials in different recording regions depends on the orientation and strength of the EEG sources, the model in Fig. 3 allows for a meaningful relative comparison of signal amplitude.

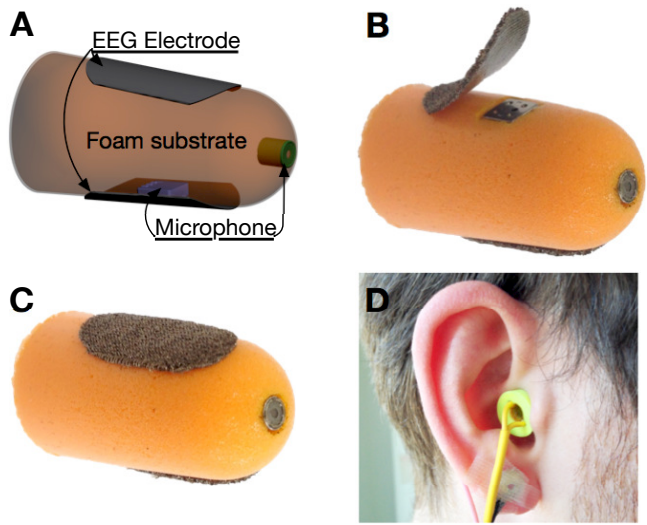

**Fig. 1.** Construction of the multimodal in-ear sensing device. (A) Detailed structure of the device, showing the placement of the microphone and the electrode on the substrate. (B) Construction of the multimodal sensor underneath one of the cloth electrodes. (C) Completed earpiece with electrodes and inward-facing microphone visible. (D) Placement of the earpiece in the user's ear.

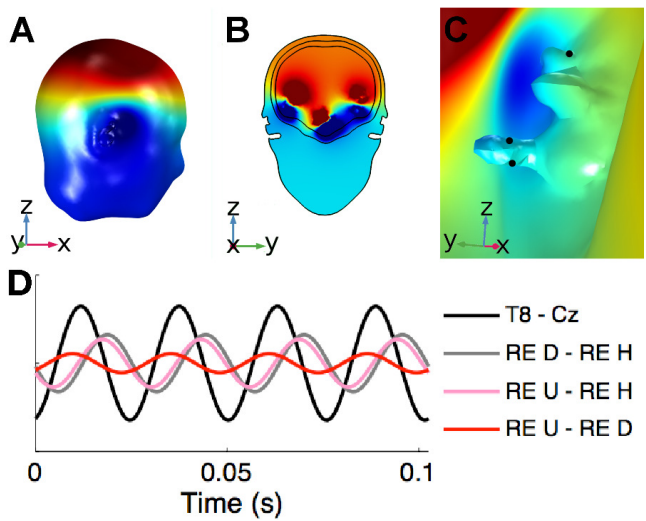

**Fig. 3.** Physics simulation in COMSOL Multiphysics® showing electric potentials on the head surface at time 0, created by dipoles oscillating at 39 Hz with positive potentials shown in red and negative potentials in blue. (A) Potentials on the whole scalp; (B) Potentials in the coronal plane; (C) Potentials in the right ear region seen from inside the head ((A), (B), and (C) have different color scales); (D) Potential differences between electrode pairs on the scalp and inside the ear canal. T8 and Cz are standard on-scalp EEG electrode positions and further abbreviations are: RE D: Right ear canal, facing downwards; RE U: Right ear canal, facing upwards; RE H: Right ear, root of the helix. The position of the ear electrodes are illustrated in black in (C).

Fig. 3D demonstrates the projected signal amplitudes from the scalp and ear canal electrodes; four full cycles of the simulated EEG dipole are shown. The potential difference was the largest between T8 and Cz, with the amplitude between the root of the helix and each of the in-ear electrodes being about half as large; the lowest amplitude was predicted between the two electrodes inside the ear canal – about a sixth of that between T8 and Cz. This clearly demonstrates the feasibility of measuring EEG using in-ear electrodes; the ear-EEG signals were expected to be weaker than those recorded from the scalp due to shorter electrode distances. The electromagnetic artifacts are also expected to be weaker inside the ear canal, as it is a cavity surrounded by conductive tissue which provides basic shielding.

**EEG acquisition from the ear canal.** Our multimodal earpiece was designed to be a comfortable and unobtrusive EEG monitor. To ensure high signal fidelity, its electrodes must provide low contact impedance with the skin over a prolonged period of time. The choice of electrode material – low impedance stretchable fabric – ensures that this requirement is satisfied even when only small amounts of saline solution are applied to the electrodes prior to insertion; the placement of the device inside the ear canal also ensures the desired low evaporation rates. Fig. 6A illustrates the stable nature of the electrode-skin impedance measured at 30 min intervals for 5 subjects over the course of their normal 8-hour working day in the office, with no restrictions placed on their activities which included having lunch and talking to people. Observe that, as desired, in all cases the median electrode impedance remained quite low, below 10 kΩ.

Having demonstrated good and stable electrode impedances, we proceeded to test the device over a number of well established EEG paradigms: ASSR, steady-state visual evoked potential (SSVEP), transient response to visual stimulus (VEP), and alpha rhythm, as summarized in Fig. 6. The ASSR at a 40 Hz modulation frequency was obtained from both ear-EEG, the mastoid (M1), and the central scalp (Cz) electrodes, and compared in Fig. 6B. Observe clear peaks at the frequency of 40 Hz for all the recording positions – the signal to noise ratio (SNR) of the EEG from the ear canal was similar to that from conventional on-scalp electrodes.

The SSVEP was induced in EEG by presenting the subjects with an LED blinking at 15 Hz. As desired, a clear peak was observed at the stimulus frequency of 15 Hz and also at its first harmonic at 30 Hz. Since the response was distributed across multiple harmonics, it was not straightforward to quantify the SNR of the recorded SSVEP, however, qualitatively it is evident that the response from the ear electrode was weaker than that of scalp electrode from a central brain region. This was expected due to a larger distance between the EEG source in the occipital region and the ear canal, as well as because of smaller electrode distances within ear-EEG.

We further demonstrated the functionality of the proposed device to acquire transient neural responses by presenting subjects with an LED switched fully ON for 200 ms and then fully OFF for 1800 ms. This response manifests itself in the negative deflection of the EEG signal approximately 180 ms after stimulus onset when recorded from the mastoid electrodes. Fig. 6D demonstrates that the waveform shape and timing of the VEP as measured from the ear electrode matched well those from scalp electrodes described in [15].

The power of alpha rhythm in EEG can be used to predict or quantify fatigue [16, 17], and is usually associated with the state of wakeful relaxation which manifests itself in the EEG oscillations in the 8 – 12 Hz frequency band, centered around 10 Hz. The loss of alpha rhythm is also one of the key features used by clinicians to define the onset of sleep. To test the capabilities of our device in measuring the alpha rhythm, the subjects were asked to sit still with the eyes open and then to close their eyes approximately 30 s into the recording. The time-frequency plot of the resulting EEG for one of the subjects is shown on Fig. 6E. The increase in the power of the alpha band is apparent after the subjects closed their eyes.

**Dealing with EEG artifacts though multimodality.** We have demonstrated in Fig. 2 that the viscoelastic earpiece, by the very nature of its substrate, helps mitigate the component of electrical noise in EEG stemming from mechanical movement of the ear canal walls against the electrode, the so called *motion artifact*. However, such noise cannot be completely suppressed by the properties of the substrate alone, particularly during strong jaw clenches, as shown in Fig. 4 (Top panel). To deal with a variety of stronger artifacts, as typically occur in real life scenarios, the proposed earpiece integrates a mechanical transducer (electret condenser microphone) within its multimodal electro-mechanical sensor, which can be used as a reference for single channel digital denoising of physiological signals [18] [19]. The results of such denoising in the jaw clenching scenario – when the microphone signal corresponding to the artifact is very accurate – are illustrated on Fig. 4A. Observe a significant reduction in the level of jaw clenching artifacts in the ear electrodes. In another, less favorable scenario (Fig. 4*B*), the proxy of the artifact measured by the ECM was compromised, but nevertheless mechanical noise was visibly attenuated.

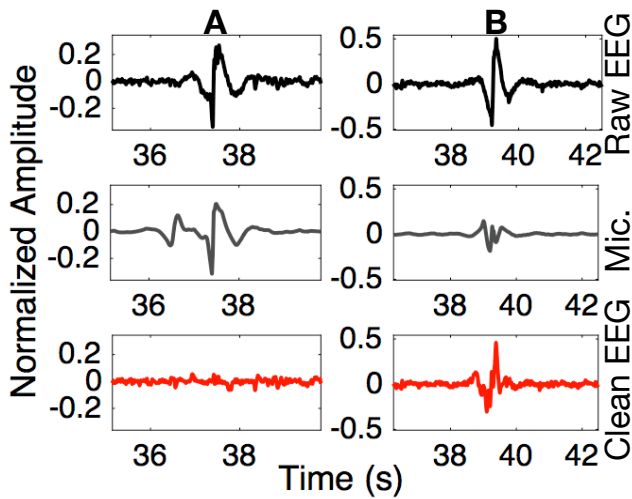

**Fig. 4.** Denoising of ear-EEG from mechanical jaw clench artifacts. (A) Best-case denoising scenario for which the artifact measurement with the embedded microphone was accurate. (B) Worst-case accuracy of artifact measurement with the embedded microphone, note that the artifact was still reduced *Clean EEG*. (Top panels) Raw EEG with a strong artifact. (Middle panels) Output of the mechanical sensor within the MMS. (Bottom panels) Denoised EEG using mechanical MMS signal as reference.

**Cardiac activity.** Another purpose of the multimodal sensor integrated within the earpiece was to sense minute mechanical movements of the tissues under the skin of the ear canal resulting from blood vessel pulsation. To verify this capability, we set up an experiment which compared the cardiac activity recorded from the proposed earpiece with the ECG obtained from the hands and PPG from a finger. Fig. 5 illustrates the recorded waveforms; observe that the integral of the pulsatile waveform from the multimodal sensor (mechanical plethysmography (MPG)) resembles the shape of the waveform from the PPG sensor. The accuracy of the measured heart rate was similar across all 3 sensors, as exemplified by the respective ECG-PPG and ECG-MPG Pearson correlation coefficients of 0.98 and 0.99. Overall, this verifies that, in the absence of jaw movements, the proposed earpiece can robustly measure the pulse rate and provide a proxy to the PPG waveform.

**Speech.** The miniature bone conduction microphone within the proposed multimodal earpiece is specifically designed for measuring acoustic signals traveling through dense tissues of the head. The original ECM transducer was modified by covering its acoustic hole with a layer of insulating material and a layer of conductive material, both of which are flexible, so that the ECM retains ability to capture acoustic signals. We next demonstrate its ability to measure speech and breathing signals when placed in close contact with the skin inside the ear canal.

In addition, to directly measure acoustic signals traveling from the vocal chords via the auditory tube, we placed an even smaller ECM at the tip of the earpiece facing towards the eardrum. The output of such a microphone would be expected to provide better speech quality than the sealed microphone within the multimodal sensor.

Subjects were asked to read out loud in their normal voice and tempo the first paragraph from the Charles Dickens' *The Tale of Two Cities*, and the resulting audio signals were simultaneously measured using: 1) an external microphone placed on the chest, 2) the multimodal sensor within the earpiece, and 3) the inward facing microphone integrated within the proposed earpiece. The speech recording waveform for one of the subjects is available in the electronic supplementary material. As expected, the quality of the speech from the external and the inward-facing microphone within the earpiece was quite similar, while the signal from the microphone within the MMS was compromised in the high frequency range but still retained intelligibility.

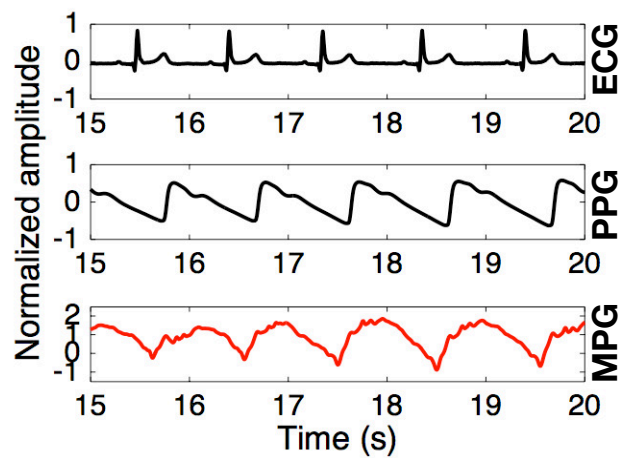

**Fig. 5.** Cardiac activity captured from inside the ear canal using the multimodal sensor embedded within the earpiece. (Top) ECG signal from the arms. (Middle) PPG signal from the finger. (Bottom) Integral of the mechanical signal produced by the microphone within the multimodal sensor.

**Respiration.** Breathing creates turbulence within the airways, such that the turbulent airflow can be measured using a microphone placed externally on the upper chest at the suprasternal notch [20]. Albeit weak, such sounds also travel through the tissues of the head and the auditory tube, so that their detection is possible even from inside the ear canal with sensitive enough microphones [21].

To demonstrate that both microphones integrated within the proposed earpiece can also measure respiration, four subjects were asked to breath in time with a metronome at different rates. At low rates (4 and 8 breaths per minute (brpm)) the acoustic signal in the ear was too low for the ECMs on the earpiece to reliably measure it. As the respiration rate increased, more turbulence was produced which increased the power of the breathing-related sound within the ear; starting from breathing rates of approximately 12 brpm we could reliably measure the breathing rates as shown in Fig. 6*F* (16 brpm) and 6*G* (28 brpm).

**Scoring sleep from ear-EEG.** We have shown so far that the proposed device is capable of measuring a variety of EEG responses – alpha rhythm, ASSR, SSVEP and VEP – as well as multiple mechanical signals associated with cardiac activity, speech and breathing. One of the key areas where such a device may find application is in monitoring sleep, assessing its quality, and possibly sleepiness. Currently, this is performed using cumbersome sleep electrodes [22].

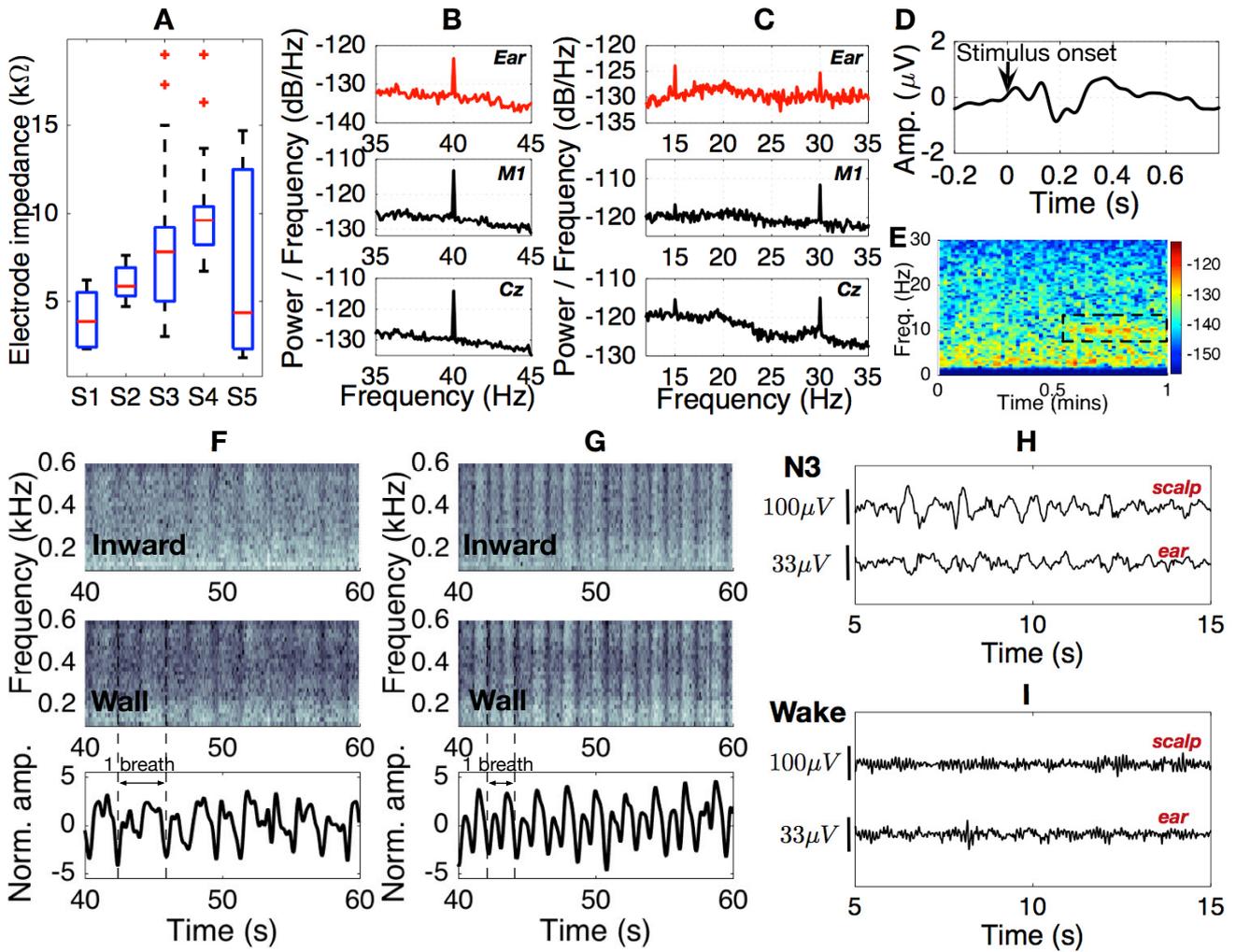

Fig. 6. EEG and respiration acquisition from inside the ear canal. (A) Stability of the electrode-skin interface over the course of 8 hours for 5 subjects. (B) ASSR response measured from the ear, M1 and Cz scalp locations. (C) SSVEP response measured from the ear, M1 and Cz locations. (D) Visual evoked potential measured with the earpiece. (E) Alpha rhythm recorded from the ear electrodes. (F) & (G) (Top and Middle) Spectrograms of the acoustic signals measured by the two microphones (inward- and wall-facing) integrated within the earpiece when the subject was asked to breath at 16 and 28 breaths per minute respectively. (F) & (G) (Bottom) Breathing signals extracted from the associated spectrograms. (H) EEG signals from scalp and ear during N3 stage of sleep. (I) EEG signals from scalp and ear when the subject was awake, but with eyes closed.

For a proof of concept of the utility of our device in sleep scoring based on EEG in particular, we recruited four subjects who were asked to reduce their sleep the night before the experiment to no more than 5 hours. On the day of the experiment the subjects slept in the afternoon for 45 minutes, with scalp and ear electrodes simultaneously recording EEG [23]. The scalp and ear-EEG signals were blinded, appropriately re-scaled, filtered and presented to sleep clinicians for scoring. Examples of a signals from the N3 and Wake stages of sleep are illustrated in Fig. 6H and 6I for scalp and ear electrodes, while Table 1 shows the scores of 360 epochs of nap EEG.

The agreement between the sleep scores from the scalp and the ear canal in distinguishing wake (Wake) from sleep (N1, N2 and N3 combined) was quantified through the kappa coefficient; its value of 0.60 (95% confidence interval 0.5 to 0.69) indicates a substantial agreement. This illustrates the promise of the proposed device in out-of-hospital assessment and diagnosis of sleep-related conditions, such as insomnia, sleep apnea, and excessive daytime sleepines.

By integrating a number of miniature sensors on an earpiece, we have demonstrated the wide-ranging potential of multimodal in-ear sensing in measuring neural, cardiac and res-piratory activity in an inconspicuous and comfortable manner. The viscoelasticity of the substrate helps mitigate the adverse effects of mechanical disturbances (motion, blood vessel pulsation) on signal quality, a critical issue in out-of-clinic EEG. Through the cross-modal information, the device can deal with even the most challenging real-life artifacts – jaw-clenches. The potential of the earpiece has been validated through case studies ranging from event related potentials and cardiac activity, through to real-life application of sleep monitoring. Future work will consider integration within the earpiece of signal acquisition, preprocessing, and wireless communication means.

|  | Wake[ear] | N1[ear] | N2[ear] | N3[ear] |
|---|---|---|---|---|
| Wake[scalp] | 70 | 19 | 17 | 0 |
| N1[scalp] | 15 | 20 | 19 | 0 |
| N2[scalp] | 7 | 17 | 120 | 9 |
| N3[scalp] | 0 | 1 | 25 | 21 |

Table 1. Contingency table of epochs for the Wake versus Sleep stages (N1, N2 and N3) for all subjects.

## Materials and Methods

**Physics simulations.** In [24] the sources of ASSR for the three amplitude-modulating frequencies, 12 Hz, 39 Hz, and 88 Hz were localized. Three source dipoles, each with two orientation components were identified (six in total): one in the brain stem with a vertical and a lateral (from the left to the right) component, and one in each auditory cortex with a tangential and a radial component. These three positions, their identified relative amplitudes and their phase onsets with respect to the stimulus were used as inputs for the current dipole sources in the model. A three-shell model with realistically shaped geometries was applied. The main shells were obtained by segmenting a magnetic resonance image (MRI) and the inner ear model was obtained from a 3D scan of an earmold. The whole model was surrounded by sphere of radius 1 m filled with air. The tissue properties were taken from the IFAC-CNR [25] based on data published in [26–28]. The model was computed using COMSOL Multiphysics® 5.2.

**Physiological measurement experiments.** In all the experiments, signal acquisition was performed using the g.tec g.USBamp amplifier with 24-bit resolution at a sampling rate of 1200 S/s. The ASSR stimulus comprised a 1 kHz tone, amplitude modulated with a 40 Hz sinusoidal signal at a 100% modulation. The numbers of volunteers involved in different measurement experiments are: SSVEP and VEP – 3, Alpha rhythm – 4, EEG denoising – 6, Speech and breathing – 4, Nap recordings – 4, Cardiac activity – 3. Sleep experiments used standard scalp electrode positions C3 and C4 referenced to A2 and A1 respectively, the ground electrode was placed on the forehead. All the experiments involving ear-EEG had the two ear electrodes referenced to a gold cup electrode placed behind the helix of the same ear, while the ground was placed on the earlobe of the same ear. All the experiments were performed at Imperial College London, under ethically approved protocol, reference ICREC 12_1_1, Joint Research Office at Imperial College London.

**Preprocessing of signals.** The ASSR and SSVEP response signals were filtered to the 1 – 45 Hz range using a 4th order Butterworth band-pass filter, with zero-phase filtering applied. The same filter was utilized for pre-processing alpha rhythm, VEP, and sleep recordings, but with 1 – 20 Hz, 1 – 13 Hz and 1 – 20 Hz bandwidths respectively. Breathing signals were filtered with a 10th order Butterworth filter to the frequency range of 100 – 600 Hz, while spectrograms used a sliding Hamming window of 0.2 s with 50% overlap.

**ACKNOWLEDGMENTS.** V.G., D.P.M., D.J.S. acknowledge support from the Rosetrees Trust. D.J.S. is supported by a National Institute for Health Research (NIHR) Professorship (NIHR-RP-011-048). D.P.M., W.v.R. and M.J.M. acknowledge support from EPSRC (EP/K025643/1).